\documentclass[sigconf,nonacm]{acmart}

\renewcommand\footnotetextcopyrightpermission[1]{}

\usepackage{algorithm}
\usepackage{algpseudocode}
\usepackage{pifont}
\usepackage{subcaption}
\usepackage{tcolorbox}
\tcbset{
    promptbox/.style={
        colback=gray!3,
        colframe=black!60,
        boxrule=0.6pt,
        arc=2pt,
        left=6pt,
        right=6pt,
        top=6pt,
        bottom=6pt,
        width=\linewidth
    }
}
\tcbuselibrary{listings, skins, breakable}

\AtBeginDocument{%
  }

\setcopyright{acmlicensed}
\copyrightyear{2018}
\acmYear{2018}
\acmDOI{XXXXXXX.XXXXXXX}
\acmISBN{978-1-4503-XXXX-X/2018/06}

\begin{document}

\title{SeDa: A Unified System for Dataset Discovery and Multi-Entity Augmented Semantic Exploration}


\author{Kan Ling}
\affiliation{%
  \institution{East China University of Science and Technology}
  \city{Shanghai}
  \country{China}
}
\email{Y30241065@mail.ecust.edu.cn}

\author{Zhen Qin}
\affiliation{%
  \institution{East China University of Science and Technology}
  \city{Shanghai}
  \country{China}
}
\email{Y40250621@mail.ecust.edu.cn}

\author{Yichi Zhu}
\affiliation{%
  \institution{East China University of Science and Technology}
  \city{Shanghai}
  \country{China}
}
\email{Y30241060@mail.ecust.edu.cn}

\author{Hengrun Zhang}
\authornote{Corresponding author}
\affiliation{%
  \institution{East China University of Science and Technology}
  \city{Shanghai}
  \country{China}
}
\email{zhanghengrun@ecust.edu.cn}

\author{Huiqun Yu}
\authornotemark[1]
\affiliation{%
  \institution{East China University of Science and Technology}
  \city{Shanghai}
  \country{China}
}
\email{yhq@ecust.edu.cn}

\author{Guisheng Fan}
\authornotemark[1]
\affiliation{%
  \institution{East China University of Science and Technology}
  \city{Shanghai}
  \country{China}
}
\email{gsfan@ecust.edu.cn}

\renewcommand{\shortauthors}{Ling et al.}


\begin{abstract}
The continuous expansion of open data platforms and research repositories has led to a fragmented dataset ecosystem, posing significant challenges for cross-source data discovery and interpretation. To address these challenges, we introduce SeDa--a unified framework for dataset discovery, semantic annotation, and multi-entity augmented navigation. SeDa integrates more than 7.6 million datasets from over 200 platforms, spanning governmental, academic, and industrial domains. The framework first performs semantic extraction and standardization to harmonize heterogeneous metadata representations. On this basis, a topic-tagging mechanism constructs an extensible tag graph that supports thematic retrieval and cross-domain association, while a provenance assurance module embedded within the annotation process continuously validates dataset sources and monitors link availability to ensure reliability and traceability. Furthermore, SeDa employs a multi-entity augmented navigation strategy that organizes datasets within a knowledge space of sites, institutions, and enterprises, enabling contextual and provenance-aware exploration beyond traditional search paradigms. Comparative experiments with popular dataset search platforms, such as ChatPD and Google Dataset Search, demonstrate that SeDa achieves superior coverage, timeliness, and traceability. Taken together, SeDa establishes a foundation for trustworthy, semantically enriched, and globally scalable dataset exploration.
\end{abstract}


\maketitle

\begin{figure*}[t]
    \centering
    \begin{subfigure}[t]{0.48\textwidth}
        \centering
        \includegraphics[width=\textwidth]{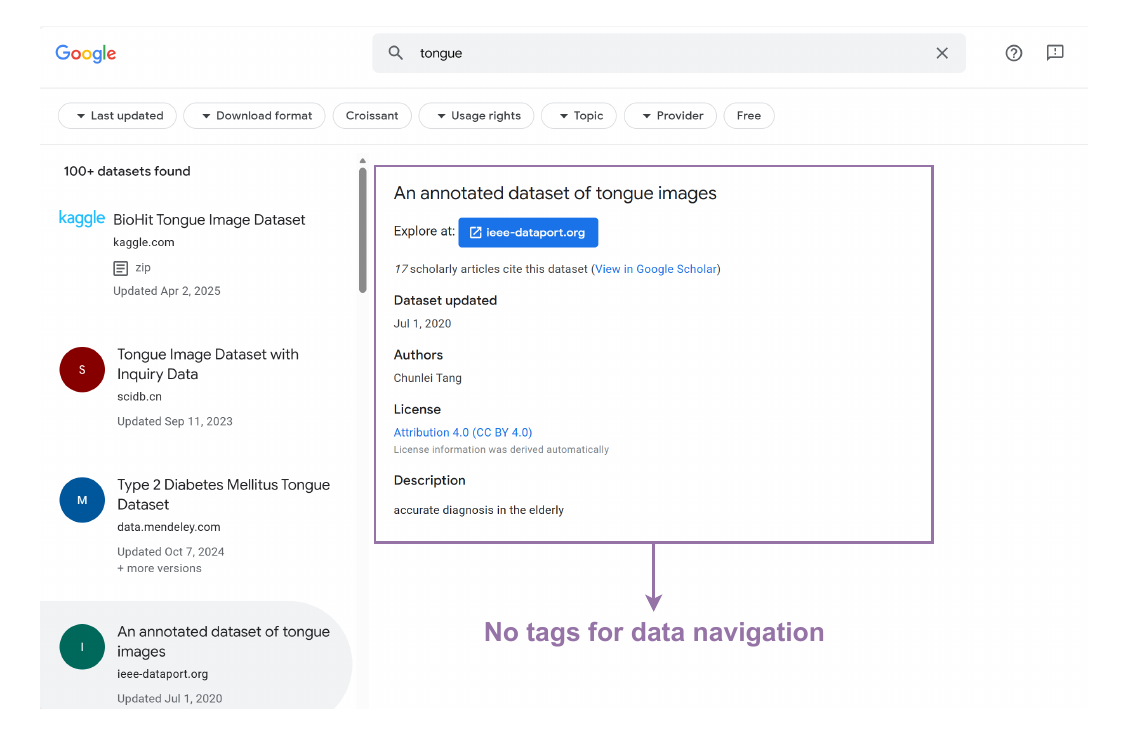}
        \caption{Google Dataset Search}
        \label{fig:google}
    \end{subfigure}
    \begin{subfigure}[t]{0.48\textwidth}
        \centering
        \includegraphics[width=\textwidth]{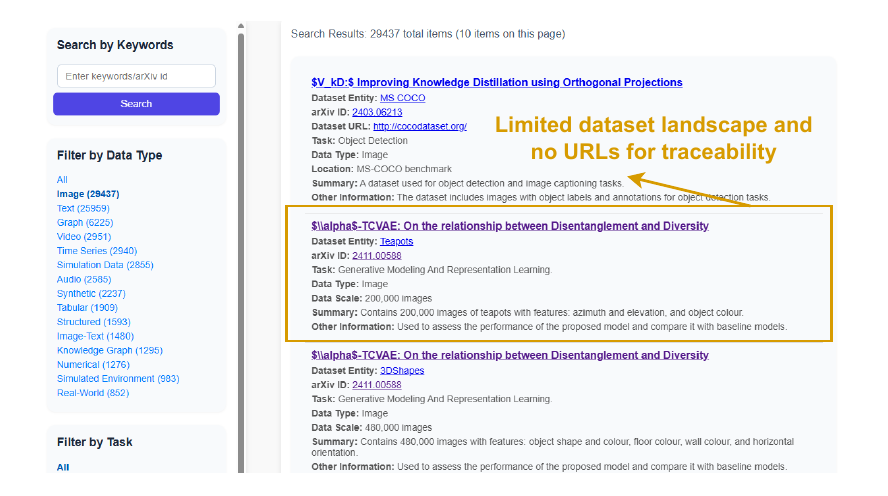}
        \caption{ChatPD}
        \label{fig:chatpd}
    \end{subfigure}
    \caption{Illustration of missing metadata elements in (a) Google Dataset Search and (b) ChatPD. Here, the dataset MS COCO has the URL, while Teapots and 3DShapes does not.}
    \label{fig:chatpd+google}
\end{figure*}

\section{Introduction} \label{sec:intro}
With the rapid advances in artificial intelligence and data science, datasets have become fundamental resources for model training, task evaluation, and scientific discovery. A growing number of open platforms are now publishing and aggregating diverse datasets\cite{attard2015systematic}, such as government-led data portals \cite{hendler2012us,data_gov_uk_2018,kassen2013promising}, academic repositories \cite{zenodo,altman2015open}, and enterprise or community-based sites \cite{abbas2021business}. However, these platforms exhibit significant heterogeneity in data formats, metadata schemas, and organizational granularity. On the other hand, users usually do not just demand for a targeted retrieval, i.e., seeking a specific dataset relevant to a particular task. Instead, they may want to systematically explore available resources. Consequently, the ways in which datasets are acquired, organized, and discovered have attracted increasing attention.

Paton et al.~\cite{paton2023dataset} identify four key components of dataset discovery and exploration: \texttt{dataset search}, \texttt{data navigation}, \texttt{data annotation}, and \texttt{schema inference}. Therein, the first component can be seen as the original goal, which can be enhanced by the latter three. The need to conduct systematic exploration centered on a research topic calls for effective data navigation mechanisms to traverse relationships between datasets. Fundamentally, addressing these user-facing challenges and obtaining a structured, unified view of the data landscape depends on foundational capabilities in data annotation and schema inference. Data annotation is required to resolve semantic inconsistencies by linking heterogeneous data elements to common vocabularies, while schema inference is necessary to generate a consolidated and abstract representation of the underlying data structures.

Despite the efforts made by current popular dataset search platforms, none of them achieves comprehensive consideration of the above four components, which restricts the applicability to some extent. For example, Google Dataset Search (GDS) \cite{brickley2019google} tackles schema inference based on \texttt{schema.org} metadata and aggregates datasets from various providers, exhibiting strong coverage. However, GDS only relies on titles or brief descriptions for search, making it insufficient for handling complex semantic queries. It also lacks effective organization and tagging mechanisms (as shown in Fig. \ref{fig:chatpd+google}(a)), limiting users' ability to quickly interpret and filter search results. On the other hand, Papers with Code (PwC, now Trending Papers in HuggingFace) \cite{martinez2021research} augments data navigation via extracting dataset references from scholarly papers and constructing paper–dataset networks to assist discovery in research contexts. Recent ChatPD \cite{xu2025chatpd} further automates data annotation to save human resources and improve efficiency. Nonetheless, due to their exclusive focus on literature sources, they exhibit limited coverage of the broader dataset landscape. Additionally, their information organization is centered around publications, often lacking access pathways (URLs) and structured representations, which hinders practical utility (as shown in Fig.\ref{fig:chatpd+google}(b)).

In this work, we propose \textbf{SeDa} (abbreviation of \textbf{Se}lect\textbf{Da}taset), a unified framework for large-scale dataset discovery and exploration. \texttt{SeDa} leverages large language models (LLMs) to conduct schema inference from not only traditional dataset metadata but also textual contents (e.g., paper abstracts, sections of full texts, or README files), greatly expanding resource coverage. Besides, three entity types are considered for the dataset repository development and data navigation augmentation, i.e., \texttt{site}, \texttt{institution}, and \texttt{enterprise}, which correspond to three major data sources: data hosting, data publishing, and data trading. Finally, \texttt{SeDa} automates data annotation \cite{limaye2010annotating,xiao2015even}, and ensures effective and non-repetitive provenance \cite{herschel2017survey,moreau2022provenance,wylot2017storing}. The relevant service has already been deployed on \textbf{https://www.selectdataset.com/}, and the key contributions are summarized as follows:
\begin{itemize}
    \item \textbf{Cross-platform standardized integration (schema inference):} We design a scalable extraction and normalization pipeline for heterogeneous data platforms with the help of LLMs, aligning field semantics and schema structures. This enables self-evolving construction of a unified dataset repository which currently contains over 7.6 million records across 200+ platforms, providing a solid foundation for cross-platform indexing and retrieval.
    \item \textbf{Multi-entity augmented retrieval (data navigation):} We introduce a dataset retrieval framework augmented by three entity types, including hosting sites, institutions, and enterprises. By integrating LLMs, the system delivers structured, coherent, and interpretable exploration results tailored to users' semantic queries.
    \item \textbf{Automated tagging and provenance (data annotation):} Beyond the core pipeline, we construct a tagging system that annotates datasets with topical domains to support efficient exploration and filtering, while relying on provenance validation to maintain traceability.
\end{itemize}

All the prompt templates used in the system pipeline are listed in Appendix \ref{app:prompt}. While the interface is currently Chinese-only, users can still search in English and rely on browser plugins for a seamless translated experience. We expect \texttt{SeDa} to ultimately serve as a comprehensive, structured, and up-to-date entry point for dataset discovery across scientific, industrial, and educational contexts.

\section{Related Work}

Data discovery has become an important topic in data management and AI communities. Existing research can be broadly categorized into three technical directions: platform aggregation and standardization systems, literature-driven and recommendation-based retrieval, and semantic enrichment and model-driven exploration. 

\textbf{Platform aggregation and standardization systems.}
Schema inference is mainly discussed in this category, with Google Dataset Search (GDS) \cite{brickley2019google} as the most representative system. It leverages \texttt{schema.org}’s dataset markup to index web pages, enabling unified access to resources from governmental, academic, and enterprise domains. This significantly improves dataset accessibility, yet the search results largely depend on surface-level metadata fields such as titles and descriptions, lacking deeper semantic organization. In parallel, the open data ecosystem has promoted interoperability standards such as W3C’s DCAT \cite{world2014data} and its European extension DCAT-AP. CKAN \cite{ckan2014ckan}, as a widely deployed open-source portal software, provides cataloging and API services for government and research institutions. Further explorations include systems such as Auctus \cite{castelo2021auctus}, which support more advanced query constraints and dataset augmentation in discovery tasks. Nevertheless, despite progress in aggregation and standardization, these systems remain limited in semantic-driven organization, multi-faceted result presentation, and cross-entity retrieval. 

\textbf{Literature-driven and recommendation-based retrieval.}
Another research line focuses on data navigation via organizing dataset information from scientific papers. Papers with Code (PwC) \cite{martinez2021research} explicitly builds a task-paper-dataset graph, greatly enhancing visibility of datasets in research contexts. Viswanathan et al. introduce DataFinder \cite{viswanathan2023datafinder}, which formalizes dataset recommendation as a natural language retrieval task. They construct a benchmark consisting of 17.5k automatically generated training queries and 392 expert-annotated test queries. Based on that, they train a bi-encoder dense retriever outperforming GDS and PwC. More recent ChatPD \cite{xu2025chatpd} leverages LLMs to automatically extract and align dataset entities from papers, enabling large-scale construction of paper-dataset networks. While these methods progressively strengthen dataset discovery in research contexts, they are inherently constrained by literature platforms and remain centered around papers, lacking cross-platform and multi-entity perspectives. 

\textbf{Semantic enrichment and model-driven exploration.}
The third line of research investigates semantic enrichment or data annotation in broader data understanding tasks. Doduo \cite{suhara2022annotating} proposes a dual-channel pretraining framework that integrates column semantics and table structure, improving column annotation and relation prediction, and thereby supporting downstream data integration and retrieval. Chorus \cite{kayali2023chorus} further demonstrates the potential of foundation models in data exploration by unifying tasks such as column type recognition, table classification, and join prediction. LLM4Tag~\cite{tang2025llm4tag} further explores graph-conditioned LLM tagging by propagating semantics over content--tag graphs and consolidating candidate topics via LLM reasoning. Though these approaches do not directly address dataset search, they highlight the potential of semantic modeling and LLMs in enhancing data discovery.

\section{Problem Definition}
Our proposed system comprehensively considers four key components in dataset discovery and exploration, which correspond to the following four problems.

\subsection{Dataset Search}
Dataset search provides a basic retrieval step that maps a user query to a set of related candidate datasets in the repository.  
Given a query \( q \), the system returns an initial result set \( R(q) \):
\[
R(q) = \mathrm{Search}(q).
\]

\subsection{Data Navigation}
Data navigation extends the initial retrieval results \( R(q) \) by incorporating topic tags (or labels) and provenance information to surface related datasets and source entities. For each dataset \( d \in R(q) \), the system searches the global collection \( D \) for datasets that share the same source (\texttt{source\_name}) or overlapping topic tags (\texttt{tag}):
\[
\begin{aligned}
D_{\text{nav}} = \{\, d' \in D \mid 
   & d'.\texttt{source\_name} = d.\texttt{source\_name} \\
   & \lor\ d'.\texttt{tag} \cap d.\texttt{tag} \neq \emptyset \,\}.
\end{aligned}
\]

In parallel, the system queries a knowledge space \(\mathcal{K} = \{S, I, E\}\) to retrieve information about the corresponding source entity, including sites ($S$), institutions ($I$), and enterprises ($E$):
\[
I_{\text{src}} = \mathrm{SourceInfo}(d.\texttt{source\_name}, \mathcal{K}).
\]

The resulting related datasets and source entities are summarized using an LLM to provide a concise overview for users:
\[
R'(q) = \mathrm{LLM\_Summarize}(D_{\text{nav}}, I_{\text{src}}).
\]

\subsection{Data Annotation}
Data annotation aims to automatically generate domain-level topic tags for each dataset, thereby improving its semantic interpretability and organizational structure.  
For each dataset, the system generates two topic tags (\texttt{tag}), corresponding to its primary research domain and application area. 
Let $D = \{ d_1, d_2, \ldots, d_n \}$ denote the dataset collection, and let $\mathcal{T} = \{ t_1, t_2, \ldots, t_m \}$ denote the vocabulary of candidate topic tags, which may expand over time (Section~\ref{subsec:tag}).
The annotation process is defined as:
\[
\mathrm{Annotate}(d_i) \rightarrow A_i, \quad A_i \subseteq \mathcal{T},
\]
where $A_i$ denotes the set of labels assigned to dataset $d_i$.

During annotation, provenance information is maintained through periodic, site-level dead-link detection, which validates dataset URLs and ensures that assigned labels correspond to accessible and traceable data sources.

\subsection{Schema Inference}
Schema inference provides a normalization mechanism that converts raw dataset descriptions from heterogeneous platforms into a unified structural representation.
For each data source $s_i$, the system applies a source-specific extraction and cleaning process, denoted as $\Lambda_{s_i}$, to produce a structured dataset object $d_i$:
\[
\Lambda_{s_i} : \text{raw description} \rightarrow d_i, \quad d_i \in D,
\]
where each $d_i \in D$ follows a standardized schema, including fields such as name, description, URL, source site, data type, and scale.

After schema normalization, redundant or near-duplicate records are removed via deduplication to yield the final dataset collection:
\[
\Gamma(D) = \{\, d_i \in D \mid \nexists d_j \in D,\, d_j \neq d_i \land \mathrm{sim}(d_i, d_j) \ge \theta \,\}.
\]
where $\mathrm{sim}(\cdot)$ denotes a similarity function over dataset representations and $\theta$ is a predefined threshold.
This process produces a clean, consistent repository for downstream retrieval and analysis.

\section{System Design} 
\begin{figure*}[t]
    \centering
    \includegraphics[width=1\textwidth]{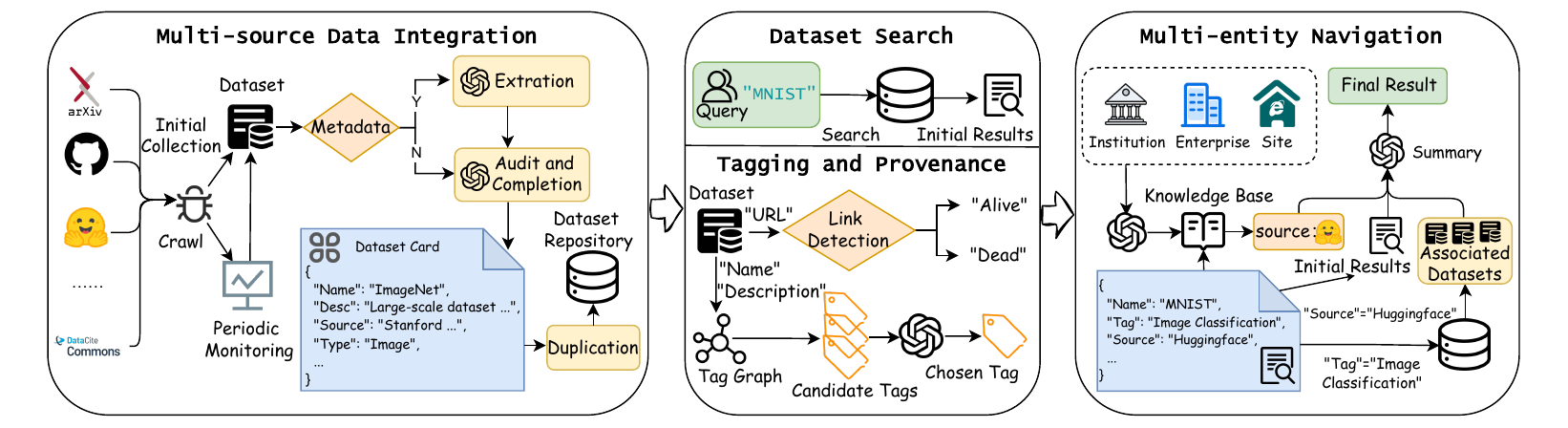}
    \caption{Overall architecture of the proposed system.}
    \label{fig:framework}
\end{figure*}

The general architecture of \texttt{SeDa} is shown in Fig. \ref{fig:framework}. We introduce three modules, i.e., multi-source data integration (for schema inference), topic tagging and provenance (for data annotation), and multi-entity augmented navigation (for data navigation), to enhance the dataset search ability.

\subsection{Multi-source Data Integration} \label{subsec:schema}
Schema inference addresses the structural and semantic heterogeneity that arises when integrating datasets from diverse repositories, a challenge widely observed in existing dataset platforms~\cite{koesten2017trials,gregory2019lost}.

\subsubsection{Data Acquisition}
Data crawling is the initial stage of our framework, aiming to collect dataset metadata from heterogeneous platforms to support subsequent processing and retrieval.

\textbf{Data Sources.}
\texttt{SeDa} integrates widely used data platforms spanning scientific research, artificial intelligence, and public governance.
Representative examples include HuggingFace~\cite{huggingface_2025} and Kaggle~\cite{kaggle_2025} for machine learning datasets, DataCite Commons~\cite{datacitecommons} and Mendeley Data~\cite{mendeleydata} for multidisciplinary scientific data, and data.europa~\cite{data_europa_2025} for governance-related open data. Community-driven platforms such as DataONE~\cite{dataone}, GBIF~\cite{gbif_2025}, and the Global Health Data Exchange contribute domain-specific datasets.

To further enhance coverage and timeliness, \texttt{SeDa} extracts dataset mentions from scholarly articles, following prior work such as ChatPD. We continuously monitor newly published arXiv~\cite{arxiv_platform} papers on a weekly basis, focusing on those containing keywords such as \texttt{dataset} and \texttt{benchmark}.
Extracted dataset mentions are treated as candidate signals and cross-validated against existing multi-source repositories. Since most datasets referenced in scholarly articles are already hosted on integrated platforms, this workflow reduces the latency between academic exposure and dataset integration.

\textbf{Crawling Methods and Update Strategies.}
To accommodate heterogeneous platforms, the framework separates data acquisition from update strategies.
Specifically, \textit{API-based acquisition} is adopted for platforms, where structured metadata can be collected via standardized APIs.
For platforms without API support, \textit{site-based crawling} is employed to perform targeted web crawling and parsing.
In addition, \textit{one-time acquisition} is used for relatively static platforms to reduce maintenance overhead, while \textit{periodic updates} are applied to frequently changing platforms to capture newly released or updated datasets. This design enables robustness and scalability across heterogeneous platforms.

\textbf{Common Crawl Supplement.}
We further incorporate Common Crawl as a supplementary data source to capture long-tail datasets that are often absent from conventional repositories. Common Crawl provides large-scale, monthly web crawl data covering a wide range of publicly accessible web content~\cite{wenzek2019ccnet}.
Within the system, Common Crawl data are processed by extracting dataset-related metadata from JSON-LD objects embedded in web pages. Fields conforming to the \texttt{schema.org} dataset specification will be parsed directly, while LLMs are used to assist in extracting and validating key attributes such as dataset name, description, and provider. These complementary strategies enable robust extraction, after which the extracted records undergo cleaning, normalization, deduplication, and update verification before integration.

\subsubsection{Data Processing}
\label{subsec:data process}
After acquisition, we process heterogeneous raw records to unify, enrich, and standardize dataset metadata, providing a consistent foundation for retrieval and navigation.

For records that already conform to the \texttt{schema.org} dataset specification, the system performs metadata validation and enrichment to ensure alignment with the unified target schema.
As highlighted by Koesten et al.~\cite{koesten2020everything}, descriptive summaries are essential for potential use cases. Accordingly, when such summaries are missing or incomplete, the system leverages LLMs to generate concise descriptions, improving semantic interpretability and retrievability.

In contrast, many heterogeneous sources lack standardized meta-data and exhibit diverse content structures. 
Following automated metadata creation principles~\cite{chapman2020dataset} and recent evidence on LLM-based extraction from unstructured text~\cite{alyafeai2025mole,xu2025chatpd}, we adopt a continuous metadata processing strategy that extracts relevant textual content and applies LLM-assisted metadata extraction during ongoing maintenance. 
For scholarly repositories such as arXiv~\cite{arxiv_platform}, abstracts and full-text sections are analyzed, while for code hosting platforms such as GitHub~\cite{github_2025}, \texttt{README} files are processed, enabling incremental updates of metadata fields and extraction strategies as new source characteristics emerge. The unified, extensible extraction prompt is provided in Appendix~\ref{subsec:appendix-extraction-prompt}.

\subsubsection{Deduplication}
Large-scale aggregation across heterogeneous platforms inevitably introduces substantial redundancy, as the same dataset is often re-published or cross-referenced with inconsistent metadata. Accordingly, \texttt{SeDa} incorporates a scalable, multi-stage deduplication pipeline that operates in three stages.

First, explicit identifier matching efficiently removes exact and near-exact duplicates by merging records with identical normalized dataset names or canonicalized URLs. 
Second, a hash-based blocking strategy prunes the candidate space, ensuring that only a small subset of potentially related records are compared further. 
Finally, semantic similarity matching is applied within each block to identify duplicates exhibiting lexical or structural variation across platforms. 
All detected duplicate relations are organized into a deduplication graph, where each connected component corresponds to a unique dataset entity. 
Within each cluster, \texttt{SeDa} selects a canonical record based on metadata completeness and source reliability, while preserving complementary attributes for provenance and traceability. 
Implementation details and cases are provided in Appendix~\ref{appendix:dedup}.

\subsection{Topic Tagging and Provenance} \label{subsec:tag}
Dataset annotation enriches records with domain-level topic tags for semantic grouping, while incorporating dead-link detection to ensure dataset availability and traceability.
\subsubsection{Topic Tag}

\begin{figure}[t]
    \centering
    \includegraphics[width=\linewidth]{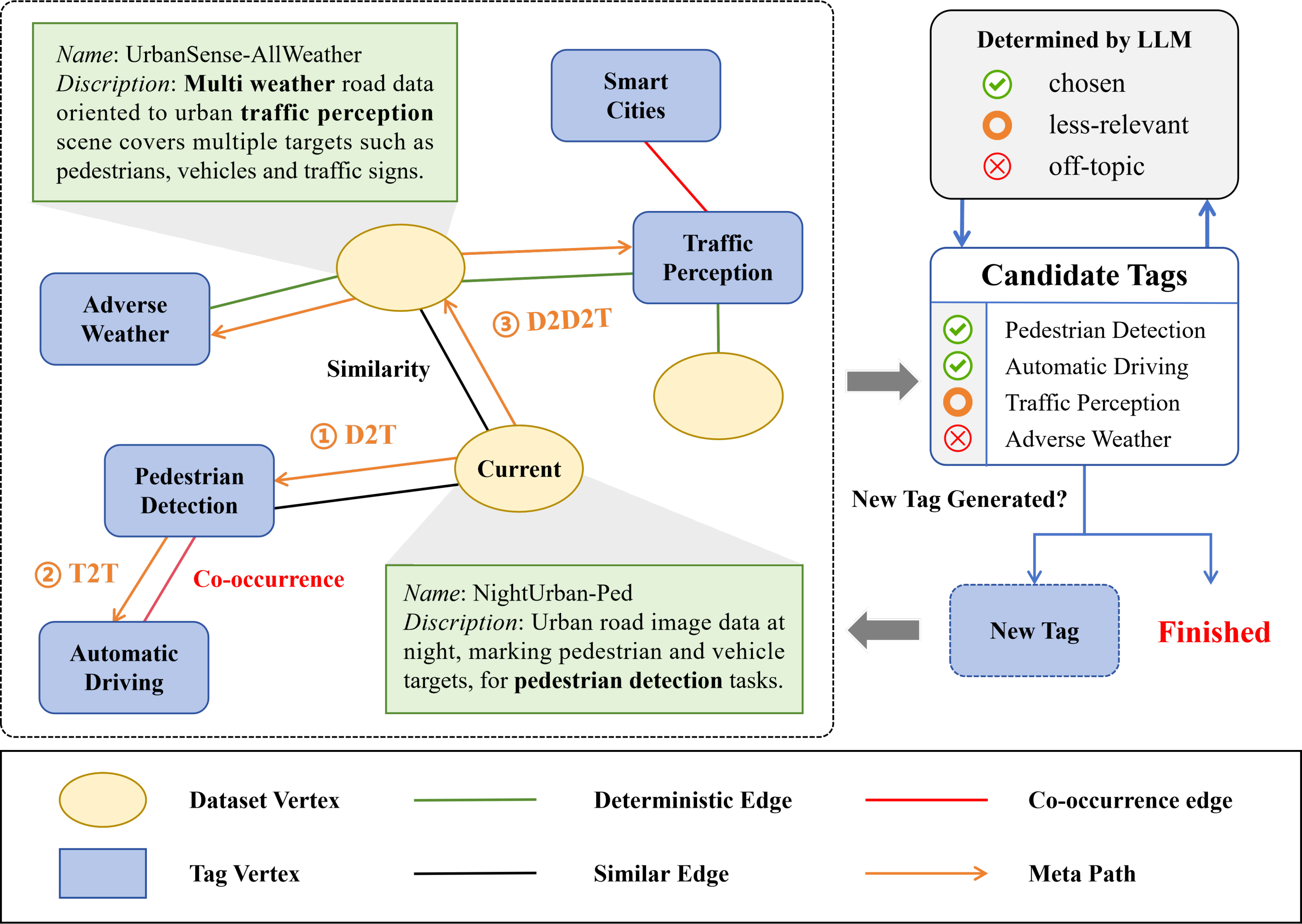}
    \caption{Tag graph construction.}
    \label{fig:tag_graph}
\end{figure}

We adapt the graph-conditioned LLM tagging system in \cite{tang2025llm4tag} (LLM4Tag) to dataset-level topic annotation, with several key modifications to better accommodate heterogeneous datasets and sparse metadata. Specifically, we augment the original content--tag recall framework with (i) an explicit 
candidate tag extraction stage to construct a compact and domain-aware initial tag pool, (ii) an additional tag--tag (T2T) relation induced by tag co-occurrence to complement the dataset-driven recall paths, and (iii) a vocabulary evolution mechanism that writes newly generated tags back into the controlled tag pool.

\textbf{Candidate Tag Pool Construction.}
For a typical industrial information retrieval system, topical annotations are often massive yet sparse \cite{tang2025llm4tag}, motivating the explicit construction of an initial tag pool. We first identify datasets with high-quality existing tags as seed sources. In parallel, for well-curated or canonical datasets with rich metadata, we employ LLMs to generate concise topical descriptions. After manual verification and quality control, these tags are normalized and deduplicated to form a compact candidate tag pool, which serves as the basis for subsequent graph construction and annotation.

\textbf{Graph Construction.}
As illustrated in Fig.~\ref{fig:tag_graph}, following LLM4Tag, each dataset is connected to candidate tags via semantic similarity, enabling dataset-to-tag (D2T, analogous to C2T in LLM4Tag) recall, while similar datasets further support multi-hop dataset-to-dataset-to-tag (D2D2T) recall. Beyond these dataset-driven paths, we introduce an additional tag-to-tag (T2T) relation derived from tag co-occurrence statistics across dataset descriptions, where edge weights reflect co-occurrence frequency~\cite{song2011automatic,cattuto2008semantic}. This extra meta-path captures latent topical associations between tags and complements the original recall graph.

\begin{figure}[t]
    \centering
    \includegraphics[width=0.95\linewidth]{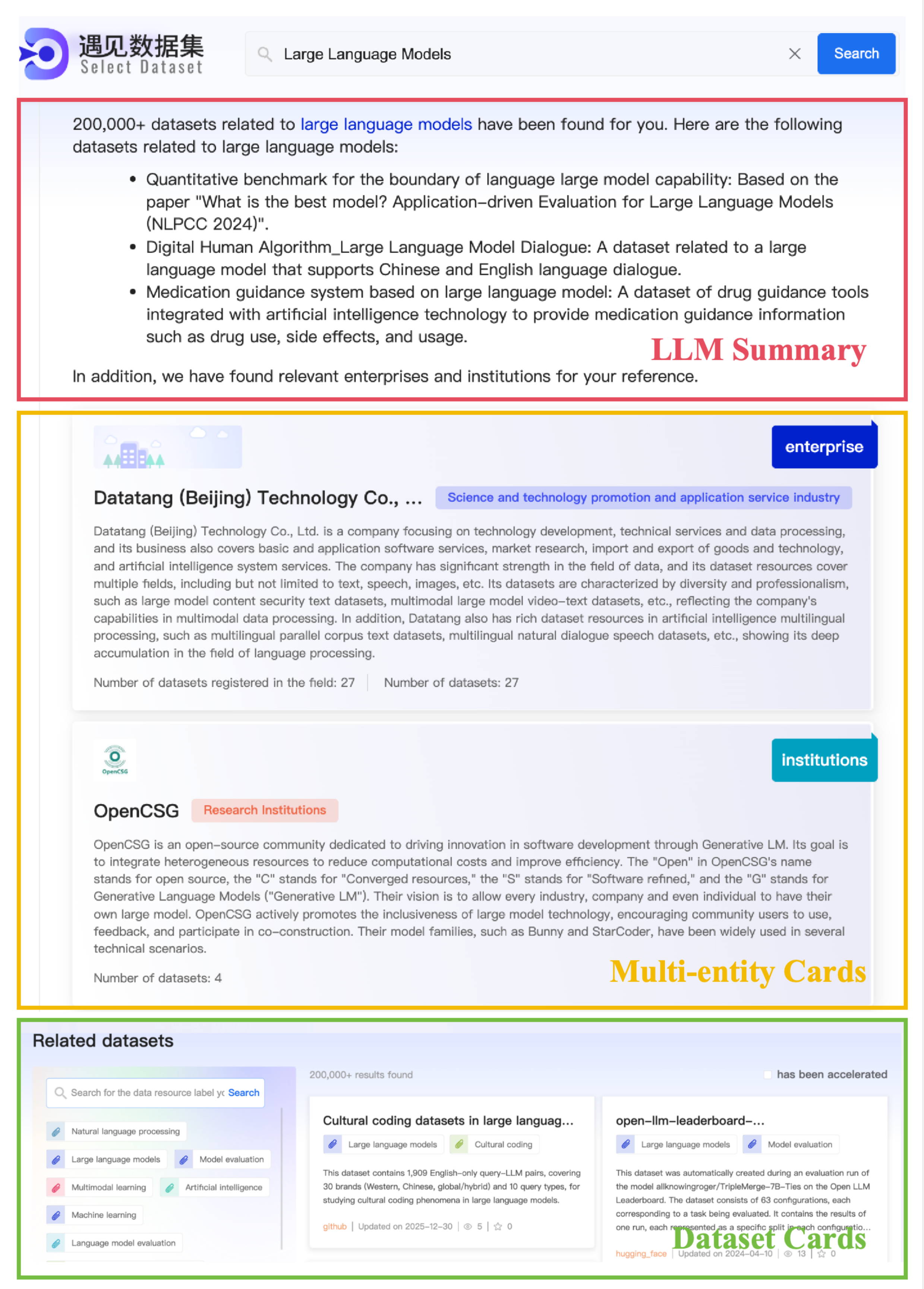}
    \caption{Navigation module visualization.}
    \label{fig:navigationUI}
\end{figure}

\textbf{Candidate Recall and Tag Refinement.}
Each dataset is represented by jointly encoding its title and description into a unified sentence embedding. 
Given a target dataset, candidate tags are recalled via the aforementioned meta-paths, yielding a compact set of potentially relevant topics (four candidates in the example shown in Fig.~\ref{fig:tag_graph}). 
Building upon this recalled set, we employ an LLM-based semantic consolidation step, following the general paradigm of LLM4Tag, to filter out irrelevant candidates and select two concise and representative tags for each dataset. In this example, the LLM discards the off-topic tag \emph{Adverse Weather} and selects \emph{Automatic Driving} and \emph{Pedestrian Detection} as the final topics for annotation. We further define a less-relevant tag \emph{Traffic Perception}, which has weak semantic relevance to but the same domain as the target dataset. This type of tags will not be used for dataset association but listed in the tag column to the left of the dataset cards (see Fig. \ref{fig:navigationUI}), which supports exploration. 

When the candidate topic tags recalled through the three graph-based paths are insufficient, this may be due to weak semantic relevance to the target dataset or the emergence of a previously uncovered domain.
Since each dataset must be annotated with two representative topic tags, we allow the LLM to supplement more suitable tags based on the dataset title and description when the recalled candidates fail to meet this requirement.
As illustrated on the right side of Fig.~\ref{fig:tag_graph}, if the LLM introduces no new tags, the annotation process terminates after updating the graph structures associated with the current dataset node. Otherwise, newly generated tags are written back into the controlled vocabulary and inserted into the tag graph, with corresponding tag–tag relations updated, enabling continuous evolution of the topic pool.

Implementation details are in Algorithm~\ref{alg:tag_generation} of Appendix~\ref{sec:appendix-topic}.

\subsubsection{Dead Link Detection}
Given the pervasive phenomenon of link decay in open data ecosystems, where a substantial fraction of dataset URLs become inaccessible over time \cite{chapekis2024online,briney2024measuring}, we design a periodic site-level dead link detection mechanism. This mechanism enforces URL verifiability at the data collection stage, ensuring that every dataset record is associated with a traceable and inspectable source, thereby preserving provenance integrity from the outset.

Different from checking individual URLs in isolation, the proposed mechanism operates at the site level and performs stratified sampling under a unified inspection budget, enabling scalable and continuous monitoring across heterogeneous data sources while avoiding excessive network and computational overhead~\cite{pan2023data,rajabi2014analyzing}.
Each week, the system samples a subset of dataset URLs associated with each site and verifies their accessibility through lightweight HTTP requests; URLs returning normal responses are regarded as \texttt{alive}, whereas those resulting in error responses, request timeouts, or domain resolution failures are classified as \texttt{dead}.
Based on these observations, the system computes the \texttt{Link Alive Rate} for each site.
When a site’s observed survival rate falls below a predefined threshold, it is flagged as a degraded source and prioritized for subsequent rechecking, and its associated datasets are temporarily withheld from the search and display modules to prevent users from accessing datasets with unreliable provenance.

To allocate inspection resources adaptively across sites of different scales and dynamics, the system assigns each site $s$ a composite importance weight $w_s$ that captures three complementary factors: site scale, historical stability, and recent update activity.
Formally, the weight is defined as: 
\[
w_s \propto N_s \cdot \sigma_s^2 \cdot \Delta N_s
\]
where $N_s$ denotes the number of indexed datasets under site $s$, $\sigma_s^2$ characterizes the temporal variability of its link survival rate, and $\Delta N_s$ reflects recent changes in dataset volume.
This formulation allows the monitoring policy to jointly account for coverage, volatility, and freshness without relying on uniform sampling.

Given a global inspection budget $K_{\text{total}}$, the sampling quota for each site is determined through normalized allocation:
\[
k_s = \frac{w_s}{\sum_{s' \in \mathcal{S}} w_{s'}} \times K_{\text{total}}.
\]

This strategy prioritizes large or unstable sources while preserving baseline coverage for smaller, stable sites under limited budgets. The complete workflow and implementation-level parameters are provided in Algorithm~\ref{alg:deadlink} of Appendix~\ref{sec:appendix-deadlink}.

\subsection{Multi-entity Augmented Navigation}
The data navigation module acts as an intermediate enhancement layer between initial retrieval and LLM-based summarization.
Without modifying the retrieval process, it augments preliminary results \(R(q)\) with dataset-level associations and source-level knowledge, which are subsequently fed into an LLM for structured summarization and presentation.

\textbf{Dataset-level Association Discovery.}
In this stage, we first perform an initial retrieval over dataset titles and descriptions using \texttt{Elasticsearch}, obtaining a result set \( R(q) \).
The relevance between a query \( q \) and a dataset record \( d \) is computed by the \texttt{BM25} ranking function \cite{robertson1994some}: 
\[
\text{Score}(q, d) = \sum_{t \in q} IDF(t) \cdot 
\frac{f(t, d) \cdot (\kappa + 1)}
{f(t, d) + \kappa \cdot \left(1 - \beta + \beta \cdot \frac{|d|}{avgdl}\right)},
\]
where \( f(t, d) \) denotes the frequency of term \( t \) in the representation of dataset \( d \), and \( IDF(t) \) is the inverse document frequency. The remaining parameters $\kappa$ and $\beta$ are standard BM25 hyperparameters controlling term-frequency saturation and document-length normalization, respectively, while $|d|$ and $avgdl$ denote the length of $d$ and the average document length over the corpus.

We then perform dataset-level association discovery to identify related datasets. 
Datasets are grouped into a contextual neighborhood \( D_{\text{nav}} \) based on shared \texttt{source\_name} and overlapping \texttt{tag} attributes, which capture provenance-based and thematic relationships and further enrich the original query results.
Each associated dataset is presented as a structured card to support exploration of both datasets and their sources.

\textbf{Source-level Knowledge Enrichment.}
Each \texttt{source\_name} is mapped to a corresponding site, institution, or enterprise entity and queried against the associated knowledge base.
The retrieved entity information includes descriptive metadata such as name, background, domain, and activity focus. These attributes are organized into entity-level knowledge cards and integrated into the navigation interface, enabling the navigation module to expose dataset ecosystems from complementary perspectives spanning hosting platforms, institutional producers, and commercial providers.

\textbf{Result Aggregation and Summarization.}
The system merges the enhanced results \( N = \{D_{\text{nav}}, I_{\text{src}}\} \) with the original retrieval output \( R(q) \).
This composite information is then provided to an LLM for contextual summarization.  
The output forms the basis of the final presentation layer, where results are organized into a unified visualization interface.

\textbf{Front-end Visualization.}
As shown in Fig.~\ref{fig:navigationUI}, the visualization interface adopts a three-tier layout.  
At the top, an LLM-generated summary provides a concise overview of the queried topic and highlights representative datasets.  
The middle section presents multi-entity cards, including institutions and enterprises, each displaying the entity name, brief description, and domain information, enabling users to explore relevant institutions, follow institutional dataset releases, or identify potential commercial opportunities.  
The bottom section shows related dataset cards, allowing users to directly access datasets and their corresponding platforms for further exploration.

\section{Experiment}
In this section, we first present large-scale dataset integration results (Section~\ref{subsec:schema1}) to demonstrate the robustness of schema inference, next conduct user-query--centric ablation (Section~\ref{subsec:ablation}) to quantify the contribution of individual components under realistic retrieval scenarios, and finally perform platform-level comparisons (Section~\ref{subsec:plat}) to evaluate \texttt{SeDa}’s effectiveness in emerging dataset discovery and dataset exploration.

\subsection{Dataset Integration Results} \label{subsec:schema1}

\textbf{Overall Integration Results.}
We report the overall integration performance of the framework described in Section \ref{subsec:schema} on large-scale heterogeneous datasets. To date, the framework has integrated a total of \textbf{7,683,121 dataset records} from more than two hundred platforms. DataCite Commons \cite{datacitecommons} accounts for over 2.1 million records, data.europa \cite{data_europa_2025} for more than 1.76 million, DataONE \cite{dataone} for nearly 0.96 million, and Mendeley Data \cite{mendeleydata} for approximately 0.45 million. Machine learning platforms such as HuggingFace \cite{huggingface_2025} and Kaggle \cite{kaggle_2025} each contribute several hundred thousand records. In addition, hundreds of thousands of records are aggregated from governmental open data portals and research institution repositories. After deduplication, approximately \textbf{600,000 redundant records} are merged, resulting in a unified entity-level dataset index.

\textbf{Common Crawl Supplement.}
As a supplementary source, Common Crawl identified approximately 750,000 candidate records, of which about 700,000 were incorporated after rule-based filtering and consistency verification. In addition, 2,907 distinct domains were detected, including 662 domains hosting at least 100 datasets and 76 domains hosting more than 1,000 datasets. These high-frequency sources provide signals for continuous monitoring and incremental integration.

These results validate the effectiveness of the framework in large-scale multi-source dataset integration and support subsequent retrieval and discovery.

\subsection{User-query--centric Ablation} \label{subsec:ablation}
To evaluate the necessity of individual components in \texttt{SeDa} under realistic query scenarios, we conduct a user-query--centric ablation study based on anonymized natural language dataset search queries collected from our deployed prototype system.
These queries reflect genuine dataset discovery needs and are collected independently of the evaluated modules to avoid evaluation bias. Each query is expressed in free-form natural language, exhibiting diverse topic scope, granularity, and intent complexity.
Depending on the interaction pattern under evaluation, different subsets of these queries are identified to reflect the functionality of each module.
All system variants operate on the same indexed corpus and differ only in whether topic-aware filtering, dead-link detection, or multi-entity exploration is enabled, 
with retrieved results analyzed from complementary perspectives including topic relevance, link availability, and exploration effectiveness. The quantitative ablation results are summarized in Table~\ref{tab:ablation}.

\begin{table}[t]
\setlength{\tabcolsep}{3.5pt}
\caption{Query-centric ablation results.}
\label{tab:ablation}
\centering
\small
\begin{tabular}{l|cc|c|c}
\hline
Variant 
& \multicolumn{2}{c|}{Topic Relevance} 
& Availability 
& Exploration \\
& Win\% 
& Score 
& Alive\%  
& Datasets (Gain\%) \\
\hline
Full (SeDa)          & 84.40 & 2.60 & 99.84 & 519 (+10.9\%) \\
w/o Topic           & 10.14 & 1.67 & --    & --         \\
w/o Dead-link       & --    & --   & 94.68 & --         \\
w/o Multi-entity    & --    & --   & --    & 468        \\
\hline
\end{tabular}
\end{table}

\begin{table}[t]
\caption{LLM evaluation breakdown.}
\label{tab:topic_llm_detail}
\centering
\setlength{\tabcolsep}{3.5pt}
\begin{tabular}{l|ccc|ccc}
\hline
Evaluator 
& \multicolumn{3}{c|}{Preference (\%)} 
& \multicolumn{3}{c}{Relevance Score (0--3)} \\
& Direct & Topic & Tie 
& Direct & Topic & $\Delta$ \\
\hline
GPT-5.2        & 9.38  & 84.38 & 6.24 & 1.615 & 2.619 & +1.004 \\
Gemini~3~Pro   & 10.90 & 85.20 & 3.90 & 1.689 & 2.607 & +0.918 \\
DeepSeek-V3.2  & 11.70 & 83.60 & 4.70 & 1.629 & 2.539 & +0.910 \\
Qwen3-Max      & 8.60  & 84.40 & 7.00 & 1.739 & 2.642 & +0.903 \\
\hline
Avg.           & 10.14 & 84.40 & 5.46 & 1.668 & 2.602 & +0.934 \\
\hline
\end{tabular}
\end{table}

\begin{table*}[tb]
\caption{Availability and metadata of representative newly discovered datasets (June--September 2025).}
\begin{center}
\begin{tabular}{|l|c|c|c|l|l|}
\hline
\textbf{Dataset Name} & \textbf{Ours} & \textbf{ChatPD} & \textbf{GDS} & \textbf{Tag} & \textbf{Source} \\
\hline
SDSS-V DR19 \cite{pallathadka2025double} & \ding{51} & \ding{55} & \ding{55} & astronomy, stellar observation & github \\
\hline
LOONGBENCH \cite{huang2025loong} & \ding{51} & \ding{55} & \ding{55} & reasoning enhancement, synthetic data generation & github\\
\hline
GlobalBuildingAtlas \cite{zhu2025globalbuildingatlas} & \ding{51} & \ding{55} & \ding{55} & geospatial data, sustainable development & mediaTUM \\
\hline
AthleticsPose \cite{suzuki2025athleticspose} & \ding{51} & \ding{55} & \ding{55} & sports data analysis, human pose estimation & github \\
\hline
AgriPotential \cite{sakka2025agripotential} & \ding{51} & \ding{55} & \ding{55} & agricultural potential prediction, remote sensing & zenodo \\
\hline
\end{tabular}
\label{tab:case_comparison}
\end{center}
\end{table*}

\textbf{Topic-aware Refinement.}
We evaluate the effectiveness of \texttt{SeDa}’s topic-aware refinement for fine-grained dataset discovery and retrieval enhancement.
In practical scenarios, users often start with a coarse topic and progressively refine their needs toward specific subtopics (e.g., ``cloud computing'' $\rightarrow$ ``cloud security''). Here, the relationship between the coarse topic and specific subtopics can be either focused or exploratory. To capture this interaction pattern, we track user topic-refinement behaviors in the deployed system and extract the corresponding anonymized queries from search logs. The resulting query set spans over 50 high-level topics, each associated with at least two subtopics, yielding approximately 180 decomposable user queries.

To assess the contribution of topic-aware refinement, we consider two retrieval settings for each such user intent:
(i) directly retrieving datasets using the combined query, and 
(ii) retrieving datasets under the main topic followed by topic-aware filtering toward the subtopic.
For each query pair, we construct Result A and Result B by selecting the top-10 retrieved datasets from direct retrieval and topic-aware filtered retrieval, respectively.
We then evaluate subtopic relevance using the prompt described in Appendix~\ref{subsec:appendix-topic-prompt}, with GPT-5.2, Gemini~3~Pro, DeepSeek-V3.2, and Qwen3-Max serving as independent evaluators.
Each model assigns relevance scores on a unified 0--3 scale, and the final relevance to the subtopic is aggregated across evaluators, who are blind to the retrieval strategy or the correspondence between results and methods.
The aggregated scores reported in Table~\ref{tab:ablation} (Topic Relevance) are averaged across all evaluators, while detailed per-model results are provided in Table~\ref{tab:topic_llm_detail}.

In addition, to qualitatively assess the quality of the generated topic annotations, we compare our generated tags for several representative datasets with the tags provided by HuggingFace.
The comparison results are presented in Appendix~\ref{subsec:appendix-topicexp}.
The results indicate that our topic annotations are more fine-grained and semantically informative, and that topic-aware refinement consistently yields higher subtopic relevance than direct retrieval, confirming that \texttt{SeDa}’s topic annotation module effectively enhances retrieval quality.

\textbf{Dead-link Detection.}
We evaluate the impact of dead-link detection by disabling the availability filtering module and comparing the accessibility of retrieved datasets under identical user queries.
This evaluation is conducted over approximately 300 anonymized real user queries sampled from the deployed system, reflecting realistic dataset discovery workloads.
Using the link availability assessment strategy described in Section~4, we compare the percentage of accessible dataset links with and without dead-link detection, observing that disabling this module yields substantially more invalid links, while enabling it consistently improves alive ratios across queries (see Availability Alive in Table~\ref{tab:ablation}).

Beyond this query-level ablation, we conduct two rounds of site-level inspections, with consistently higher alive rates observed in the second round, confirming the effectiveness of the detection mechanism. Detailed sampling formulation and results are reported in Appendix~\ref{sec:appendix-deadlink}.

\textbf{Multi-entity Exploration.}
We evaluate the effectiveness of \texttt{SeDa}’s multi-entity exploration mechanism for large-scale dataset discovery. This evaluation is conducted over the same set used for dead-link detection.
In practical search scenarios, users often explore related entities such as hosting platforms, institutions, or companies to uncover additional resources beyond the initial query results.
To quantify this effect, we compare discovery outcomes with and without multi-entity exploration enabled.
For each query, we record the number of datasets returned by initial retrieval, then enable entity-based navigation and measure the additional datasets discovered through associated entities.
The exploration gain is defined as the percentage increase relative to the initial results.
Results (Exploration Datasets in Table~\ref{tab:ablation}) show that multi-entity exploration consistently yields substantial relative gains in discoverable datasets, demonstrating that \texttt{SeDa}’s multi-entity modeling effectively expands the searchable space and enhances dataset discovery beyond query-level retrieval.

\textbf{Limitation.}
Our evaluation relies on anonymized real-world query logs collected from a deployed prototype system, reflecting realistic dataset discovery and exploration behaviors.
Nevertheless, such logs inevitably capture only a subset of possible user intents, and may not exhaustively represent rare, emerging, or highly specialized dataset search scenarios. 
We believe more comprehensive insights can be achieved with the popularity of \texttt{SeDa}.

\begin{figure}[t]
    \centering
    \includegraphics[width=0.95\linewidth]{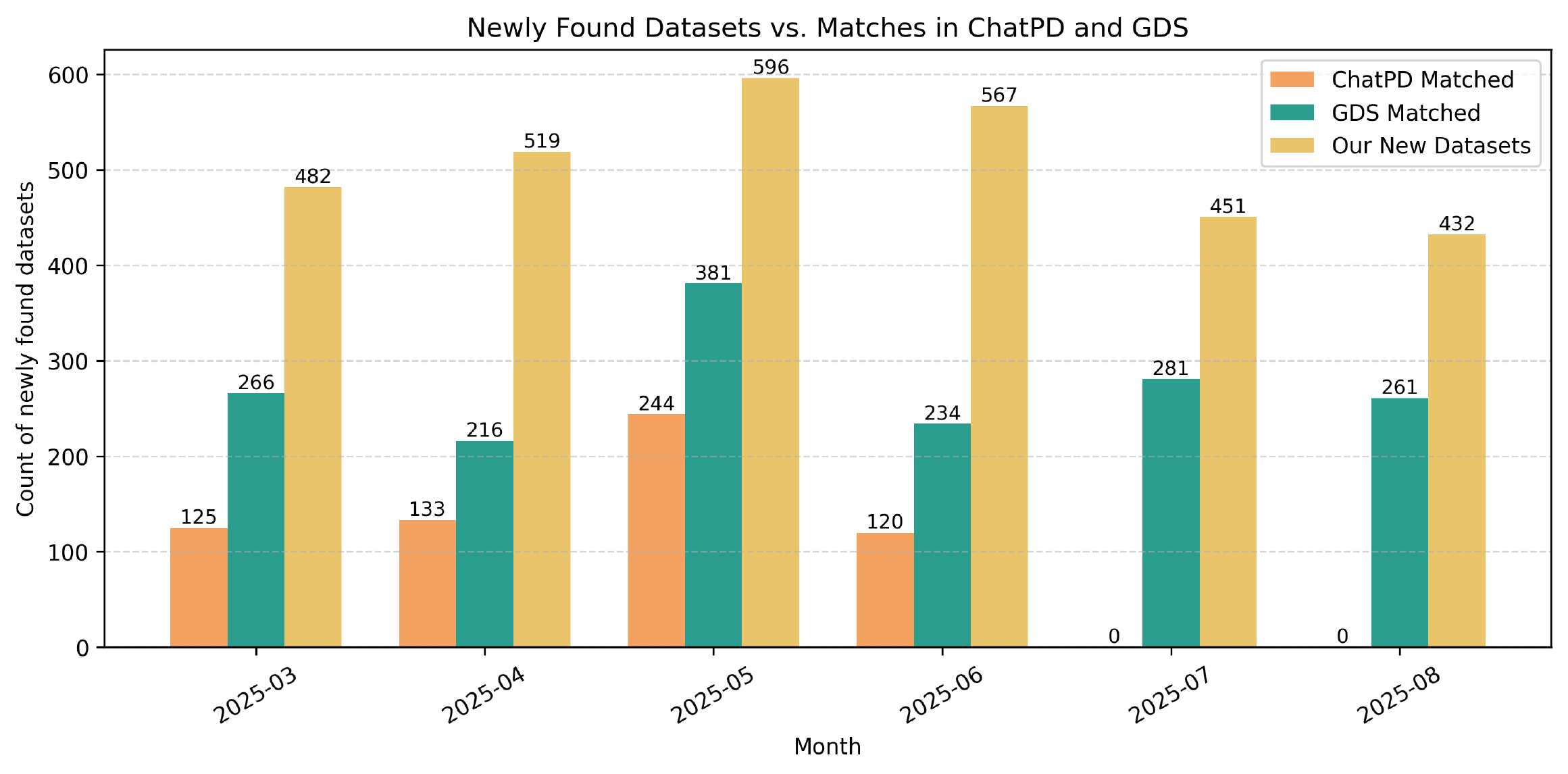}
    \caption{Monthly comparison of newly found datasets and matches in ChatPD and GDS.}
    \label{fig:new_datasets_counts}
\end{figure}

\subsection{Platform-level Comparison}\label{subsec:plat}

To evaluate the effectiveness of our framework in discovering newly emerging datasets, we examine records collected between March and August 2025 and conduct a comparative study against ChatPD and GDS.

Our system identifies 3,047 new dataset records during this period, whereas ChatPD contains only 622 and reports no corresponding entries for datasets released in July and August. GDS retrieves 1,639 datasets but still misses the remaining 1,408, indicating that even global-scale metadata search engines exhibit noticeable delays in indexing newly released resources. The monthly comparison is shown in Fig.~\ref{fig:new_datasets_counts}.

These gaps stem from fundamental differences in discovery mechanisms. ChatPD relies on datasets being cited in scholarly publications, while GDS depends on providers explicitly exposing \texttt{schema.org}-compliant metadata, rendering both pipelines inherently reactive. In contrast, our framework adopts a proactive discovery loop that continuously monitors heterogeneous platforms and incorporates long-tail sources, enabling datasets to be captured at their point of first appearance rather than waiting for downstream signals, thereby substantially improving both timeliness and coverage.

As shown in Table~\ref{tab:case_comparison}, among the 1,408 newly identified datasets, our system uncovers multiple high-impact frontier resources that are absent from both ChatPD and GDS, including milestone and first-of-the-kind datasets such as SDSS-V DR19 \cite{pallathadka2025double}, LOONGBENCH \cite{huang2025loong}, GlobalBuildingAtlas \cite{zhu2025globalbuildingatlas}, AthleticsPose \cite{suzuki2025athleticspose}, and AgriPotential \cite{sakka2025agripotential}. These datasets span large-scale astronomical surveys, state-of-the-art LLM benchmarks, globally consistent building-level atlases, open 3D athletic pose benchmarks, and multispectral remote sensing resources for agricultural potential prediction.

Collectively, these results demonstrate that \texttt{SeDa} achieves both high-quality discovery and strong timeliness by identifying datasets at their first appearance and bridging the gap between academic outputs and data infrastructure.

\section{Conclusion}

In this work, we developed \textbf{SeDa}, a unified framework for large-scale dataset discovery and exploration. The system integrates heterogeneous data from over two hundred platforms through schema standardization, tag annotation, and multi-entity navigation. By combining topic tags for semantic organization with multi-entity navigation over enterprises, institutions, and sites, \texttt{SeDa} extends dataset retrieval into a structured knowledge space, enabling cross-platform exploration with improved consistency, interpretability, and provenance control. The resulting repository of more than 7.6 million standardized records demonstrates the scalability and practical utility of our approach. 

While \texttt{SeDa} currently focuses on metadata-level representations for dataset discovery and organization, we plan to further enhance it with capabilities for table-level understanding and task-oriented retrieval~\cite{hulsebos2024took}, enabling the framework to interpret the internal structure of datasets and align them more closely with analytical or application needs. Such extensions will further transform \texttt{SeDa} from a comprehensive dataset repository into an intelligent environment for knowledge-rich and context-aware data exploration.

\bibliographystyle{ACM-Reference-Format}
\bibliography{main}

\appendix

\section{Topic Annotation}
\label{sec:appendix-topic}

\subsection{Algorithm}
\label{subsec:appendix-topicalg}
The following Algorithm~\ref{alg:tag_generation} provides pseudocode for the topic tag generation and association module, complementing the methodology in the main text.

\begin{algorithm}[htbp]
\caption{Topic Annotation}
\label{alg:tag_generation}
\begin{algorithmic}[1]
\Statex \textbf{Input:} A dataset collection $\mathcal{D}$ (each $d=\langle t_d, x_d, \rangle$ with title $t_d$, description $x_d$);
\Statex \textbf{Output:} Final topic tags $\{\mathcal{T}_d\}_{d\in\mathcal{D}}$ and an updated controlled vocabulary $\mathcal{V}$ with an evolving tag graph $\mathcal{G}$.

\Statex \textbf{Stage I: Candidate Tag Pool Construction}
\State Select seed datasets $\mathcal{D}_{seed}\subseteq\mathcal{D}$ with reliable existing tags.
\State $\mathcal{V} \leftarrow \bigcup_{d\in\mathcal{D}_{seed}} \mathrm{Tags}(d)$
\State Select canonical datasets $\mathcal{D}_{canon}\subseteq\mathcal{D}$ with rich metadata.
\ForAll{$d\in\mathcal{D}_{canon}$}
    \State $\mathcal{V} \leftarrow \mathcal{V} \cup \textsc{LLMGenerateTopics}(t_d, x_d)$
\EndFor
\State merge duplicates tags for $\mathcal{V}$ Define $\mathcal{D}_{init} \leftarrow \mathcal{D}_{seed} \cup \mathcal{D}_{canon}$.

\vspace{2pt}
\Statex \textbf{Stage II: Graph Construction}
\State obtain semantic embeddings and Initialize graph $\mathcal{G}$
\ForAll{$d \in \mathcal{D}_{init}$, $v \in \mathcal{V}$} \State $e_d \gets \mathrm{Emb}([t_d;\,x_d]),\; e_v \gets \mathrm{Emb}(v)$ \EndFor

\ForAll{$d \in \mathcal{D}_{init}$}
    \ForAll{$v \in \mathrm{Tags}(d)$}
        \State Add D2T edge $(d \rightarrow v)$ to graph $\mathcal{G}$
    \EndFor
    \ForAll{$d' \in \mathcal{D}_{init}\setminus\{d\}$}
        \If{$\mathrm{sim}(e_d, e_{d'}) \ge \delta$}
            \State Add D2D edge $(d \rightarrow d')$ to graph $\mathcal{G}$
        \EndIf
    \EndFor
\EndFor

\ForAll{$(v_i, v_j) \in \mathcal{V} \times \mathcal{V}$ with $\mathrm{co\text{-}occurrence}(v_i,v_j)>\tau$}
    \State Add T2T edge $(v_i \leftrightarrow v_j)$ to graph $\mathcal{G}$
\EndFor

\vspace{2pt}
\Statex \textbf{Stage III: Candidate Recall and LLM Selection}
\ForAll{$d \in \mathcal{D}$}
    \State Find similar datasets $\mathcal{D}'_d$ and related tags $\mathcal{V}'_d$ in $\mathcal{G}$
    \State \textsc{Candidates}$\mathcal{C} \gets \textsc{Recall}_{\mathrm{D2T},\,\mathrm{D2D2T},\,\mathrm{T2T}}(d,\mathcal{D}'_d,\mathcal{V}'_d)$
    \State $\mathcal{T}_d\ \gets \textsc{LLMSelect}(t_d, x_d, \mathcal{C})$
    \If{\textsc{LLMSelect introduces new tags}}
        \State Add such tags into $\mathcal{V}$ and update $\mathcal{G}$
    \EndIf
\EndFor

\State \Return $\{\mathcal{T}_d\}_{d\in\mathcal{D}}$, $\mathcal{V}$, and $\mathcal{G}$.
\end{algorithmic}
\end{algorithm}

\subsection{Topic Quality}
\label{subsec:appendix-topicexp}
To qualitatively evaluate our topic tag generation, we compare our tags with those provided by HuggingFace~\cite{huggingface_2025} on representative datasets, as shown in Table~\ref{tab:tag_case_study}. While HuggingFace typically provides generic tags such as \texttt{audio classification} or \texttt{image classification}, 
\texttt{SeDa} can identify more fine-grained and domain-specific topics such as 
\texttt{traditional Chinese instruments} and \texttt{tree species classification}. These results indicate that platform-provided tags often omit key contextual or culturally specific information, whereas our approach can capture such nuanced semantics and enhance the interpretability of dataset metadata.

\begin{table*}[t]
\centering
\caption{Comparison between our generated topic tags and HuggingFace annotations.}
\label{tab:tag_case_study}
\begin{tabular}{|p{1.6cm}|p{4cm}|p{2.5cm}|p{3cm}|p{5cm}|}
\hline
\textbf{Dataset} & \textbf{Our Tags} & \textbf{HuggingFace Task} & \textbf{HuggingFace Tag} & \textbf{Description (excerpt)} \\ 
\hline
\textbf{CTI5} & traditional Chinese instruments, music classification & audio classification & music, art & Recordings of Chinese traditional musical instruments for timbre recognition. \\ 
\hline
\textbf{Rico} & mobile application, user interface design & others & graphic-design & Large-scale dataset of mobile app screenshots and UI metadata. \\ 
\hline
\textbf{TreeSatAI-T} & tree species classification, temporal series analysis & image classification & earth-observation, remote-sensing & TreeSatAI time series dataset for tree species classification. \\ 
\hline
\textbf{batman2} & traditional Chinese medicine, drug discovery & others & biology, drug discovery & Molecular structures derived from traditional Chinese medicine for drug discovery. \\ 
\hline
\textbf{clinc\_oos} & intent classification, out-of-scope query detection & text classification & intent classification & Dataset for out-of-scope intent detection in dialogue systems. \\ 
\hline
\end{tabular}
\end{table*}

\section{Deduplication Implementation Details and Examples}
\label{appendix:dedup}

\begin{table*}[t]
\centering
\caption{Example of duplicate records detected by semantic similarity where explicit identifier matching fails.}
\label{tab:tinyimagenet}
\begin{tabular}{|p{2.5cm}|p{7cm}|p{7cm}|}
\hline
\textbf{Field} & \textbf{Record 1} & \textbf{Record 2} \\
\hline
\textbf{dataset\_name} & Tiny\_Imagenet & Tiny ImageNet \\
\hline
\textbf{dataset\_desc} & Tiny Imagenet has 200 Classes, each class has 500 traininig images, 50 Validation Images and 50 test images. Label Classes and Bounding Boxes are provided. & Tiny ImageNet contains 100000 images of 200 classes (500 for each class) downsized to 64×64 colored images. Each class has 500 training images, 50 validation images and 50 test images. \\
\hline
\textbf{dataset\_url} & https://figshare.com/articles/dataset/Tiny\_Imagenet & https://paperswithcode.com/dataset/tiny-imagenet \\
\hline
\end{tabular}
\end{table*}
This section provides a concise yet comprehensive description of the three-stage deduplication pipeline implemented in \texttt{SeDa}.

In the first stage, explicit identifier matching is used to detect cross-platform replicas by aligning normalized dataset names and canonicalized URLs.
Next, to avoid the quadratic complexity of global pairwise comparison, a hash-based blocking strategy is employed to partition the candidate space.
Dataset titles and descriptions are concatenated into a unified textual representation and converted into text signatures.
A hybrid blocking approach~\cite{papadakis2020blocking,moslemi2024evaluating} combining SimHash~\cite{charikar2002similarity} and Locality-Sensitive Hashing (LSH)~\cite{indyk1998approximate} is adopted: SimHash captures character-level similarity, while LSH preserves cosine similarity under TF--IDF~\cite{salton1988term} vectorization.
Only records that fall within the same hash bucket are considered potential duplicates and forwarded to the subsequent semantic matching stage, substantially reducing the comparison cost.

In the semantic similarity assessment stage, \texttt{dataset\_name} and
\texttt{dataset\_desc} are concatenated to form a unified text string for
each record and encoded into an embedding using the sentence transformer
\texttt{all-MiniLM-L6-v2}.
Cosine similarity between dataset embeddings is computed, and record pairs whose similarity exceeds a predefined threshold are considered duplicates and merged. Semantic similarity search within each hash bucket is implemented using an approximate nearest neighbor index~\cite{muja2009fast}, with FAISS~\cite{johnson2019billion} employed to ensure scalable and efficient retrieval.

All detected duplicate relations are organized into a deduplication graph, where each connected component corresponds to a unique dataset entity.
Within each component, \texttt{SeDa} selects a canonical record to represent the entity using a rule-based strategy that accounts for metadata completeness, source reliability, and descriptive richness.
Authoritative sources with more comprehensive metadata are preferred, while auxiliary information such as alternative URLs and attributes from other records is retained to preserve provenance traceability.

Table~\ref{tab:tinyimagenet} presents a representative example illustrating the necessity of semantic similarity matching.
The two records correspond to the \texttt{Tiny ImageNet} dataset but originate from different platforms with distinct URLs and slightly different naming conventions.
Although explicit identifier matching and hash-based blocking alone are insufficient to resolve this duplication, semantic similarity matching correctly identifies the two records as duplicates. In this example, the cosine similarity between the two semantic representations reaches 0.89, exceeding the similarity threshold used in our system (0.85), thereby triggering a correct merge. This example demonstrates how the full pipeline complements the high-level deduplication design described in the main paper by resolving subtle cross-platform redundancy.

\section{Adaptive Site-level Dead-link Detection}
\label{sec:appendix-deadlink}
This section presents validation and implementation details of our dead-link detection. Algorithm~\ref{alg:deadlink} outlines the site-level detection procedure, and Table~\ref{tab:deadlink_exp} summarizes inspection results across representative platforms with varying scales at two system stages.

The first inspection round was conducted on June 25, 2025, shortly after repository initialization, when large-scale multi-source integration was still ongoing.
The second round took place on September 20, 2025, after major data cleaning and consolidation had completed and the system reached a relatively stable state.
Across all examined sources, alive rates exhibit consistent improvements in the second round, approaching near-complete accessibility, which confirms the effectiveness of the proposed monitoring mechanism.

\begin{algorithm}[htbp]
\small
\caption{Adaptive Site-Level Dead-Link Detection}
\label{alg:deadlink}
\begin{algorithmic}[1]
\Require Site set $\mathcal{S}$ with statistics $\{N_s,\sigma_s^2,\Delta N_s\}$, total budget $K_{\text{total}}$, and threshold $\tau = 0.9$
\Ensure Alive-site list $\mathcal{S}_{\text{alive}}$

\State \textbf{(Weighting)}
\For{$s \in \mathcal{S}$}
  \State $w_s \gets \log(1+N_s)\,(1+\lambda_1\sigma_s^2)\!\left(1+\lambda_2\frac{\Delta N_s}{N_s+\varepsilon}\right)$
\EndFor

\State \textbf{(Budget Allocation)}
\For{$s \in \mathcal{S}$}
  \State $k_s^{\star} \gets \dfrac{w_s}{\sum_{s' \in \mathcal{S}} w_{s'}}\, K_{\text{total}}$
  \State $k_s \gets \min\!\big(k_{\max},\,\max(k_{\min},\,\operatorname{round}(k_s^{\star}))\big)$
\EndFor

\State \textbf{(Sampling and Checking)}
\For{$s \in \mathcal{S}$}
  \State Randomly sample $k_s$ URLs from site $s$
  \State Perform HTTP checks and compute $\operatorname{AliveRate}(s)$
  \If{$\operatorname{AliveRate}(s) < \tau$}
    \State Mark $s$ as \textbf{DEAD}, hide all datasets of $s$, and schedule rechecking
  \Else
    \State Mark $s$ as \textbf{ALIVE}
  \EndIf
\EndFor

\Return $\mathcal{S}_{\text{alive}}$
\end{algorithmic}
\end{algorithm}

\begin{table}[htbp]
\caption{Two-round site-level sampling and alive-rate results. 
($S_1$, $S_2$: sample size; 
$A_1$, $A_2$: alive rates.)}
\label{tab:deadlink_exp}
\begin{center}
\begin{tabular}{|l|r|r|c|r|c|}
\hline
\textbf{Site} & \textbf{$N_s$} & \textbf{$S_1$} & \textbf{$A_1$} & \textbf{$S_2$} & \textbf{$A_2$} \\
\hline
DataCite & 2{,}103{,}855 & 2{,}598 & 0.89 & 2{,}696 & \textbf{0.99} \\
\hline
data.europa & 1{,}761{,}468 & 2{,}564 & 0.88 & 2{,}654 & \textbf{1.0} \\
\hline
Kaggle & 359{,}561 & 1{,}148 & 0.96 & 1{,}199 & \textbf{1.0} \\
\hline
HuggingFace & 193{,}227 & 858 & 0.85 & 879 & \textbf{0.98} \\
\hline
data.gov & 71{,}424 & 531 & 0.86 & 534 & \textbf{1.0} \\
\hline
arXiv & 17{,}743 & 272 & 0.90 & 266 & \textbf{1.0} \\
\hline
OpenData & 7{,}758 & 180 & 0.74 & 176 & \textbf{0.99} \\
\hline
Papers with Code & 7{,}217 & 175 & 0.94 & 169 & \textbf{1.0} \\
\hline
Aliyun Tianchi & 4{,}170 & 132 & 0.96 & 129 & \textbf{1.0} \\
\hline
Shanghai Data Open & 1{,}021 & 59 & 0.79 & 63.9 & \textbf{0.98} \\
\hline
\end{tabular}
\label{tab:deadlink_exp}
\end{center}
\end{table}

\section{Prompt Templates Used in the System Pipeline} \label{app:prompt}
To provide a complete view of the system workflow, this section summarizes the prompt templates used across different stages of the pipeline. Each subsection corresponds to a specific system module and documents the associated prompt template, clarifying how textual inputs and intermediate signals are structured for interaction with large language models.

\subsection{Description Extraction Prompt}
\label{subsec:appendix-isdataset-prompt}
This subsection introduces the prompt used to extract missing dataset descriptions and improve metadata completeness. Such cases commonly occur in datasets collected through web crawling, where descriptive information is primarily presented on dataset web pages.

The prompt provides the model with the page HTML as the main information source, while the URL is included only as contextual metadata referencing the hosting platform or organization.

\begin{figure}[htb]
\centering
\begin{tcolorbox}[promptbox, title={Prompt for Dataset Description Generation}]
\small
\textbf{System} \\
You are a dataset description extraction expert. Your task is to extract and generate a descriptive text for a dataset from the given dataset detail page.

\textbf{Requirements}
\begin{itemize}
    \item Extract and generate dataset-related descriptive information based solely on the provided page content.
    \item Exclude website-level explanatory text and any content unrelated to the dataset itself.
    \item Do not include any explanations, reasoning steps, or intermediate analysis in the output.
    \item The output should be a coherent natural-language description rather than a list of metadata fields.
\end{itemize}

\textbf{Notes}
\begin{itemize}
    \item Do not introduce any information that does not appear on the dataset detail page.
    \item Ensure that the generated description is accurate, objective, and non-misleading.
    \item The provided page URL is used only as contextual metadata and must not be used to infer information beyond the page content.
\end{itemize}

\textbf{Input} \\
The dataset detail page URL is: \texttt{\{dataset\_url\}} \\
The HTML content of the dataset detail page is: \texttt{\{html\_content\}}

\end{tcolorbox}
\end{figure}

\subsection{Dataset Extraction Prompt}
\label{subsec:appendix-extraction-prompt}
This subsection presents the unified dataset-centric extraction prompt introduced in Section~\ref{subsec:data process}

In general, source adapters indicate the input text type and highlight platform-specific cues for extraction.
For GitHub README text, the adapter focuses on dataset-centric descriptions while filtering incidental mentions.
For arXiv titles and abstracts, it emphasizes signals of dataset introduction or release, prioritizing dataset naming, construction methodology, scale, and author or institutional cues when available.
For HuggingFace dataset descriptions, it guides the extraction of overall dataset descriptions, content composition, split information, and access or usage instructions.

\begin{figure}[htbp]
\centering
\begin{tcolorbox}[promptbox, title={Unified Prompt for Dataset Extraction}]
\small

\textbf{System} \\
You are a dataset analysis expert specializing in extracting dataset-related information from unstructured textual records originating from heterogeneous sources.
You will be provided with a piece of text content. Please carefully analyze the given text, and perform dataset-centric information extraction according to the instructions below.

\textbf{Source Adapter} \\
\texttt{\{source\_hint\}}

\textbf{Requirements}
\begin{itemize}
    \item Identify explicit or implicit dataset information in the text.
    \item If dataset information is present, extract and summarize its core characteristics, including the dataset name, overall description, and access or acquisition method.
    \item If no meaningful description can be obtained, leave the corresponding fields as empty strings.
\end{itemize}

\textbf{Response Format}
\begin{verbatim}
{
  "is_dataset": "Yes/No/Uncertain",
  "dataset_name": "",
  "dataset_desc": "",
  "dataset_url": "",
  "analysis": ""
}
\end{verbatim}

\textbf{Notes}
\begin{itemize}
    \item Do not access or rely on any external links in the text.
    \item All extracted information must be based on the provided content.
    \item The output should be concise and written in an academic style appropriate for dataset documentation.
    \item Do not include any comments, explanations, or formatting outside the specified JSON structure.
\end{itemize}

\textbf{Input}
\begin{itemize}
    \item Text Content: \texttt{\{record\_text\}}
\end{itemize}

\end{tcolorbox}
\end{figure}

\begin{figure*}[htbp]
\centering
\begin{tcolorbox}[promptbox, title={Prompt for Topic Filtering Evaluation}]
\small

\textbf{System} \\
You are a professional dataset relevance evaluation expert.
The user is searching for datasets related to the topic ``\{query\_text\}'', with a specific subtopic ``\{tag\_meaning\}''. Your task is to assess how well two sets of retrieved results match the user’s true needs. Result A and Result B are produced by two different retrieval strategies. Please remain neutral and base your judgment solely on dataset relevance and quality.

\vspace{4pt}
\textbf{Evaluation Criteria}
\begin{itemize}
    \item \textbf{Subtopic Alignment}: The core content of the dataset should correspond to ``\{tag\_meaning\}'', rather than merely mentioning related keywords.
    \item \textbf{Topic Consistency}: The dataset should align with ``\{query\_text\}'', rather than belonging to cross-domain or weakly related scenarios.
    \item \textbf{Semantic Relevance}: The actual data content, application context, or annotation targets should closely match the user’s intent, avoiding overly generic or ambiguous matches.
\end{itemize}

\vspace{4pt}
\textbf{Evaluation Procedure}

\textit{Step 1: Item-Level Scoring} \\
Independently score each item in Result A and Result B using integer values:

\begin{itemize}
    \item 3 (Perfect): Clearly satisfies ``\{tag\_meaning\}''.
    \item 2 (Relevant): Generally relevant, but with minor deviations, vague descriptions, or limited noise.
    \item 1 (Weak): Mainly keyword-level matches and unlikely to satisfy real user needs.
    \item 0 (Irrelevant): Largely unrelated to the user intent, or clearly not a dataset (e.g., papers, models, tools, or code repositories).
\end{itemize}

For each item, provide one extremely concise justification.

\vspace{4pt}
\textit{Step 2: List-Level Aggregation} \\
Compute:
\begin{itemize}
    \item score\_A: the average score of all items in Result A
    \item score\_B: the average score of all items in Result B
\end{itemize}

Determine the winner:
\begin{itemize}
    \item The result set with the higher score wins.If the two scores are equal, output ``Tie''.
\end{itemize}

\vspace{4pt}
\textbf{Output Format (JSON only)}
\begin{verbatim}
{
  "items_A": [
    {"rank": 1, "score": 0-3, "reason": "..."},
    ...
  ],
  "items_B": [
    {"rank": 1, "score": 0-3, "reason": "..."},
    ...
  ],
  "score_A": <float>,
  "score_B": <float>,
  "winner": "Result A" | "Result B" | "Tie",
  "reason": "A Brief explain"
}
\end{verbatim}

\vspace{4pt}
\textbf{Retrieved Results}

\begin{itemize}
    \item Result A: \texttt{\{result\_a\}}
    \item Result B: \texttt{\{result\_b\}}
\end{itemize}

\end{tcolorbox}
\end{figure*}

\subsection{Topic Filtering Evaluation Prompt}
\label{subsec:appendix-topic-prompt}
This subsection illustrates the prompt design for LLM-based topic relevance evaluation. The prompt takes a query, a subtopic definition, and two retrieval results as input, and guides the model to produce fine-grained relevance scores and a final comparison in a standardized JSON format.

\begin{figure*}[htb]
\centering
\begin{tcolorbox}[promptbox, title={Prompt for Topic Tag Refinement}]
\small
\textbf{System} \\
You are a dataset annotation specialist responsible for assigning standardized topic tags to datasets.
You will be provided with a dataset name, a textual description, and a set of candidate topic tags, and your task is to refine and finalize the dataset-level topic annotation.

\textbf{Requirements}
\begin{itemize}
    \item Select \textbf{exactly two} representative topic tags that best characterize the dataset at the dataset level.
    \item Generate new tags only when the provided candidate tags are insufficient to produce two suitable and representative tags.
    \item Classify all remaining candidate tags as either weakly related or discarded based on their semantic relevance.
\end{itemize}

\textbf{Response Format}
\textbf{Response Format}
\begin{verbatim}
{
  "selected": [
    {"tag": string, "is_new": boolean, "reason": string},
    {"tag": string, "is_new": boolean, "reason": string}
  ],
  "weakly_related": [string],
  "discarded": [string]
}
\end{verbatim}

\textbf{Notes}
\begin{itemize}
    \item \texttt{selected} must contain exactly two tags.
    \item New tags must appear in \texttt{new\_tags}.
    \item Output JSON only.
\end{itemize}

\textbf{Input}
\begin{itemize}
    \item Dataset Name: \texttt{\{dataset\_name\}}
    \item Dataset Description: \texttt{\{dataset\_description\}}
    \item Candidate Tags: \texttt{\{candidate\_tags\}}
\end{itemize}
\end{tcolorbox}
\end{figure*}

\begin{figure*}[htbp]
\centering
\begin{tcolorbox}[promptbox, title={Prompt for Query Result Summarization}]
\small

\textbf{System} \\
You are a dataset summarization specialist responsible for producing concise and
coherent summaries of datasets in response to user queries.

\textbf{Requirements}
\begin{itemize}
    \item Summarize the datasets relevant to the given user query in a neutral and
    informative manner.
    \item Dataset descriptions should be paraphrased for clarity and conciseness,
    rather than copied verbatim from the input.
    \item If provider information (e.g., institutions, enterprises, or platforms) is
    available, include a brief description; otherwise, omit this part.
\end{itemize}

\textbf{Response Format} \\
A single textual summary introducing the relevant datasets and, when applicable,
the associated provider information.

\textbf{Notes}
\begin{itemize}
    \item The response should contain only the summarized textual content.
    \item No additional explanations, markup, or formatting should be included.
\end{itemize}

\textbf{Example} \\
\emph{User Query:} Datasets related to autonomous driving \\
\emph{Dataset Records:} \\
KITTI Vision Benchmark Suite: A widely used autonomous driving dataset supporting tasks such as object detection, 3D tracking, stereo vision, and road segmentation across diverse driving scenes and conditions. \\
Cityscapes: An urban street-scene dataset designed for road and scene understanding, with images collected from multiple cities and seasons. \\
\emph{Provider Information:} \\
DeepScenario: A research organization focused on autonomous driving scenario generation to enhance perception and decision-making capabilities. \\
\emph{Output:} Representative autonomous driving datasets include the KITTI Vision Benchmark Suite for core driving perception tasks and Cityscapes for urban street-scene understanding. In addition, research organizations such as DeepScenario contribute complementary resources and expertise for autonomous driving research.

\textbf{Input}
\begin{itemize}
    \item User Query: \texttt{\{user\_query\}}
    \item Dataset Records: \texttt{\{dataset\_records\}}
    \item Provider Information (optional): \texttt{\{provider\_info\}}
\end{itemize}

\end{tcolorbox}
\end{figure*}

\subsection{Topic Refinement Prompt}
\label{subsec:appendix-refinement-prompt}
This subsection presents the prompt used for dataset-level topic tag refinement.The prompt instructs the LLM to identify exactly two representative topics for a dataset, allows the generation and explicit marking of new tags when existing candidates are insufficient, and labels the remaining tags as either less relevant or off topic based on semantic relevance.

\subsection{Summarization Prompt}
\label{subsec:appendix-summary-prompt}
This subsection presents the prompt used for dataset summarization in response to user queries.
Given a user query and a set of retrieved dataset records, the prompt instructs the LLM to produce a concise, neutral summary of relevant datasets, optionally incorporating provider information when available.

\end{document}